\documentclass[runningheads]{llncs}
\usepackage{todonotes}
\usepackage{tikz}
\usepackage[utf8]{inputenc}       \usepackage{multirow}  
\usepackage{float}     
\usepackage{array}     
\usepackage{tikz}
\usetikzlibrary{backgrounds, positioning} 
\usepackage{float}
\usepackage{algorithm}
\usepackage{algpseudocode}
\usepackage{hhline}    
\usepackage{complexity}

\usepackage{graphicx}
\usepackage{soul}
\usepackage[T1]{fontenc}             
    
\usetikzlibrary{graphs} 
\usepackage{mathtools}
\usepackage{graphicx}
\usepackage{hyperref}
\usepackage{comment}

\newcommand{\bigo}{\mathcal{O}}

\begin{document}

\title{Correlation Clustering with Overlap: a Heuristic Graph Editing Approach}

\author{Faisal N. Abu-Khzam, Lucas Isenmann and Sergio Thoumi}
\authorrunning{Abu-Khzam, Isenmann and Thoumi} 
\institute{
Department of Computer Science and Mathematics\\
Lebanese American University, 
Beirut, Lebanon.\\
\email{\{faisal.abukhzam,lucas.isenmann\}@lau.edu.lb}\\
\email{sergio.thoumi@lau.edu}
}
\maketitle 

\begin{abstract}

Correlation clustering seeks a partition of the vertex set of a given graph/network into groups of closely related, or just close enough, vertices so that elements of different groups are not close to each other. 
The problem has been previously modeled and studied as a graph editing problem, namely {\sc Cluster Editing}, which assumes that closely related data elements must be adjacent. As such, the main objective (of the {\sc Cluster Editing} problem) is to turn clusters into cliques as a way to identify them. This is to be obtained via two main edge editing operations: additions and deletions. There are two problems with the {\sc Cluster Editing} model that we seek to address in this paper. First, ``closely'' related does not necessarily mean ``directly'' related. So closeness should be measured by relatively short distance. As such, we seek to turn clusters into (sub)graphs of small diameter. Second, in real applications, a data element can belong, or have roles, in multiple groups. In some cases, without allowing data elements to belong to more than one cluster each, makes it hard to achieve any clustering via classical partition-based methods. We address this latter problem by allowing vertex cloning, also known as vertex splitting. 
Heuristic methods for the introduced problem are presented along with experimental results showing the effectiveness of the proposed model and algorithmic approach.


\end{abstract}

\section{Introduction}

Clustering aims to partition data points into different groups/clusters such that similar or closely related points are in the same cluster. Clustering has prominent usage in machine learning and other domains. For example, it can be used for pooling in Graph Neural Networks (GNNs)
\cite{bianchi2020spectralclusteringgraphneural}, training Convolutional Neural Networks (CNNs) \cite{caron2019deepclusteringunsupervisedlearning}, and Visual Categorization \cite{csurka2004visual}.
A known challenge with some classical methods such as $k$-means \cite{hartigan1979k} and min-sum \cite{10.1145/335305.335373} is the requirement to specify the number of clusters as it adds a layer of difficulty for the user and leads to inaccuracies when the chosen number of clusters is incorrect.

Bansal et al. \cite{bansal2004correlation} introduced the notion of {\em correlation clustering} where pairwise correlations between elements of a data set are represented by a graph, or network, with an edge being labeled "+" or "-" indicating whether the two corresponding elements are similar or closely correlated. Correlation clustering then seeks to maximize the number of "+" within clusters and the number of "-" between them. Notably, correlation clustering approaches do not require the user to specify the number of clusters. This paved the way to formulating clustering as a graph modification problem such as {\sc Cluster Editing} and {\sc Cluster Deletion} \cite{shamir2004cluster}. 
In these formulations, a cluster is considered to be a clique, that is a subgraph where vertices are pairwise adjacent. 
The objective is to obtain a disjoint union of cliques/clusters by adding and deleting (cluster editing) or only deleting (cluster edge deletion) at most $k$ edges. In the same work, both problems were proven to be $\NP$-Hard. Huffner et al. studied a similar clustering problem where only vertex deletion is allowed \cite{huffner2010fixed}, namely {\sc Cluster Vertex Deletion} (CVD), which is also known to be $\NP$-Hard \cite{lewis1980node}.

In some application domains, requiring a cluster to be a clique could be too strict to provide the needed information. For instance, relaxation models such as $s$-club models could capture more protein interactions in Protein-Protein Interaction (PPI) networks \cite{balasundaram2005novel}. An $s$-club is a graph where any two vertices are at distance $s$ from each other. The 2-club and 3-club models were found to be ideal relaxation models in Biological Networks \cite{balasundaram2005novel,pasupuleti2008detection}. The {\sc 2-Club Cluster Editing}, {\sc 2-Club Cluster Vertex Deletion}, and {\sc 2-Club Cluster Edge Deletion} problems  were all shown to be $\NP$-Complete \cite{liu2012editing}. In general, bounded diameter clustering was studied from a theoretical standpoint \cite{alpert1997splitting,chang2014complexity} and was shown to be effective in multiple application domains \cite{sohaee2009bounded,krishnan2006efficient}. Other endeavors studied minimizing the sum of cluster diameters instead of bounding the diameter of a single cluster \cite{hansen1987minimum,charikar2001clustering,bilo2005geometric}.

Although correlation clustering alleviated the requirement to specify the number of clusters, it still shared a common pitfall with previous models. Namely, clustering via graph modification always resulted in non-overlapping clusters, i.e. each data element was required to be in at most one cluster. In practice, a data point can be part of multiple clusters. For example, proteins can have multiple biological functions \cite{nepusz2012detecting}, so clustering a PPI network should allow for a protein to be in multiple clusters. This has led to studying clustering with overlapping communities in a graph theoretical context \cite{baumes2005finding,fellows2011graph}. In this paper, we cluster overlapping communities by splitting a vertex $v$ into two vertices whose neighborhoods form a partition of the neighborhood of $v$.
This is combined for the first time with edge-deletion into 2-clubs and can be easily generalized to $s$-clubs. The notion of vertex splitting was first introduced in the real of community detection \cite{gregory2007algorithm}. Cluster Editing with Vertex Splitting (CEVS) was introduced in \cite{abu2018cluster} and it was shown to be $\NP$-Complete in \cite{abukhzam2023cluster}.
 
\vspace{10pt}
\noindent
{\bf Our contribution.} 
We evaluate the utility of adopting 2CCEDVS as a model for correlation clustering with overlaps. In particular, we present, implement, and test a novel heuristic approach. Our presented empirical evaluation clearly show the notable effectiveness of 2CCEDVS
when compared to the best-known algorithms for correlation clustering. 

\section{Preliminaries}

We work with simple, undirected and unweighted graphs. We denote a graph by $G = (V, E)$, where $V$ and $E$ represent its sets of vertices and edges, respectively.
We use standard terminology of graph theory as presented in \cite{west2001introduction}. The open-neighborhood of a vertex $w$ is the set of vertices adjacent to it, denoted by $N(w)$. The closed neighborhood of $w$ is $N[w] = N(w) \cup w$. The degree of a vertex is the number of edges incident on it, which is the number of its neighbors $|N(w)|$ since we are dealing with simple graphs. 

\noindent 
A subgraph $G'=(V',E')$ of $G$ has $V'$ and $E'$ as subsets of $V$ and $E$, respectively. A clique is a (sub)graph consisting of pair-wise adjacent vertices. A path $P$ is a sequence of distinct vertices $(v_1, v_2, v_3,  \ldots, v_{t})$ in a graph where $v_i$ and $v_{i+1}$ are adjacent $\forall i \in \{1,\ldots, t-1\}$. 
The distance from vertex $u$ to $w$ is the shortest path between them and is denoted by $d(u, w)$. A path of length $n$ is denoted as $P_n$. A cycle is a sequence of three or more vertices $v_1,v_2,\ldots,v_n$ where $v_1=v_n$,  and $v_1,\ldots,v_{n-1}$ is a path. A cycle of length 3 is a triangle, while a cycle of length 4 is a square. A conflict triple is a path of length three. 
An $s$-club is a set of vertices any two of which are within distance $s$ from each other. 

The {\sc Cluster Edge Deletion} problem can be solved in $O(1.404^k) $\cite{tsur2022cluster} while the fastest asymptotic running-time for {\sc Cluster Editing} is in $O(1.62^k+m+n)$ \cite{bocker2012golden} where $k$ is the number of allowed modifications, $m$ is the number of edges, and $n$ is the number of vertices. The best algorithm for CEVS runs in $O(2^{7k\log k}+n+m)$ \cite{arrighi2023cluster}. CEVS is more computationally demanding due to the numerous ways one can split a single vertex. Replacing cliques with $s$-clubs might not make the corresponding problems solvable in faster asymptotic running times. However, the parameter $k$ in the respective exponential function is assumed to be much smaller. This is because one potentially needs to delete less edges to obtain a disjoint union of 2-clubs, for example, then to obtain a disjoint union of cliques. The {\sc $s$-Club Cluster Edge Deletion} problem is formally defined as follows:

\vspace{5pt}
\noindent
{\sc $s$-Club Cluster Edge Deletion} ($s$CCED)

\noindent
\textbf{Given:} A graph $G=(V,E)$, along with positive integers $s$ and $k$;

\noindent
\textbf{Question:} Can we transform $G$ into a disjoint union of $s$-clubs by performing at most $k$ edge deletions?
\vspace{5pt}

As mentioned above, when $s=2$ (i.e. 2CCED) the problem is know to be $\NP$-Hard, but $\FPT$ \cite{liu2012editing}. A trivial brute-force approach would run in $O^*(3^k)$, simply by branching on each path of length whose endpoints are at distance three from each other. Improved fixed-parameter algorithms that use more sophisticated algorithms appeared recently \cite{abu2021improved,tsur2024algorithms}. This recent development shows the increased interest in this particular problem. The current fastest fixed-parameter algorithm for 2CCED runs in $O^*(2.563^k)$ \cite{tsur2024algorithms}. 

We compare our algorithms' running time and cluster quality with the Markov Clustering (MCL) \cite{van2000graph} (C), ClusterONE \cite{nepusz2012detecting} (Java), and Karlsruhe and Potsdam Cluster Editing (KaPoCE) \cite{blasius2021pace} (C++) algorithms.
KaPoCE is the winner of the heuristic track of the 2021 PACE challenge \cite{kellerhals2021pace}.

MCL clusters a graph by simulating flow in a graph using two main operations: expansion and inflation. Expansion simply performs matrix multiplication to model the spreading out of flow. Inflation performs a Hadamard power operation followed by diagonal scaling, which shows the contraction of the flow. After these operations, each set of vertices with flow between them will be a cluster. 

ClusterONE is a greedy algorithm that aims to obtain clusters with high "cohesiveness," a measure introduced also in \cite{nepusz2012detecting}. Cohesiveness indicates how linked a vertex is to its cluster, while being separated from the rest of the graph. The algorithm starts from a seed vertex and greedily forms a cluster around it to achieve high cohesiveness. This process is repeated with different seeds, resulting in possibly overlapping clusters. At the end, highly overlapping clusters are merged and small or low-density clusters are discarded.  

KaPoCE performs Cluster Editing using the Variable Neighborhood Search (VNS) heuristic. It works by perturbing an initial solution then finding a new local optimum. This is done for multiple rounds and is combined with three refinement techniques (Label Propagation, Clique Removal and Splitting, and Vertex Swapping).

The aforementioned algorithms take different approaches to correlation clustering. MCL clusters based on network flow, ClusterONE uses cohesiveness, while KaPoCE presents a heuristic for {\sc Cluster Editing}, a graph-editing problem. Like {\sc Cluster Editing}, many models have been proposed to cluster a graph by editing it, but all share common limitations. Namely, the current models either define a cluster as a clique or restrict each vertex to a single cluster only. Defining a cluster to be a clique results in a worse clustering quality or at least an increased running-time. For example, take a clique $G$ with vertices $\{x,y,z,v\}$ and one edge $x-z$ missing; obviously, this is not considered a cluster in such models. In models where edge addition is allowed, this would increase the number of modifications needed, and thus, the execution time in practice. 
If we cannot add edges, then $\{x,y,v\}$ and $\{z\}$ cannot be in the same cluster. Isolating or deleting $z$ would increase the number of modifications and potentially reduce the clustering quality. On the other hand, if a cluster was defined to be an $s$-Club with $s\geq 2$ then this graph would not need to be modified, resulting in a faster execution-time. It would also result in an arguably better cluster since all four vertices are closely related.

Another limitation of the current graph-editing approaches is not allowing clusters to overlap, which is considered unrealistic in most application domains. Vertex splitting has been introduced as a solution, but it has been only used along with {\sc Cluster Editing} and not $s$-Club models. In this paper, we consider the introduction of vertex splitting to {\sc s-Club Cluster Edge Deletion}. By introducing vertex splitting as an additional allowed operation, we hope to combine the advantage of having a better problem model and that of requiring a smaller number of editing operations, which eventually results in more efficient algorithms. The {\sc $s$-Club Cluster Edge Deletion with Vertex Splitting} problem is formally defined as follows:

\vspace{5pt}
\noindent
{\sc $s$-Club Cluster Edge Deletion with Vertex Splitting} ($s$CCEDVS)

\noindent\textbf{Given:} A graph $G=(V,E)$, along with positive integers $s$ and $k$;

\noindent
\textbf{Question:} Can we transform $G$ into a disjoint union of $s$-clubs by performing at most a total of $k$ edge deletion and/or vertex splitting operations?

\begin{figure}
    \centering

\begin{tikzpicture}[
    scale=0.5, 
    node_style/.style={
        circle,
        draw,
        minimum size=0.5cm, 
        inner sep=0pt
    }
]

\node[node_style] (V) at (0,0) {$V$};

\node[node_style] (B3) at (0,4) {$B_3$};
\node[node_style] (B1) at (-2,2) {$B_1$};
\node[node_style] (B2) at (2,2) {$B_2$};

\node[node_style] (T3) at (-6,0) {$T_3$};
\node[node_style] (T1) at (-4,1) {$T_1$};
\node[node_style] (T2) at (-4,-1) {$T_2$};

\node[node_style] (F1) at (4,1) {$F_1$};
\node[node_style] (F2) at (4,-1) {$F_2$};
\node[node_style] (F3) at (6,0) {$F_3$};

\draw (V) -- (B1);
\draw (V) -- (B2);
\draw (B1) -- (B3);
\draw (B2) -- (B3);
\draw (B2) -- (F1);
\draw (V) -- (T1);
\draw (V) -- (T2);
\draw (T1) -- (T3);
\draw (T2) -- (T3);

\draw (V) -- (F1);
\draw (V) -- (F2);
\draw (F1) -- (F3);
\draw (F2) -- (F3);

\end{tikzpicture}
\begin{tikzpicture}[
    scale=0.5, 
    node_style/.style={
        circle,
        draw,
        minimum size=0.5cm, 
        inner sep=0pt
    }
]

\node[node_style] (V) at (0,0) {$V$};

\node[node_style] (B3) at (0,4) {$B_3$};
\node[node_style] (B1) at (-2,2) {$B_1$};
\node[node_style] (B2) at (2,2) {$B_2$};

\node[node_style] (T3) at (-6,0) {$T_3$};
\node[node_style] (T1) at (-4,1) {$T_1$};
\node[node_style] (T2) at (-4,-1) {$T_2$};

\node[node_style] (F1) at (4,1) {$F_1$};
\node[node_style] (F2) at (4,-1) {$F_2$};
\node[node_style] (F3) at (6,0) {$F_3$};

\draw (V) -- (B2);
\draw (B1) -- (B3);

\draw (B2) -- (F1);

\draw (T1) -- (T3);
\draw (T2) -- (T3);

\draw (V) -- (F1);
\draw (V) -- (F2);
\draw (F1) -- (F3);
\draw (F2) -- (F3);

\end{tikzpicture}
\begin{tikzpicture}[
    scale=0.5, 
    every node/.style={
        draw,
        circle,
        minimum size=0.5cm, 
        inner sep=0pt
    }
]
    \node (T3) at (-4,0) {$T_3$};
    \node (T1) at (-2,1) {$T_1$};
    \node (T2) at (-2,-1) {$T_2$};
    \node (V2) at (0,0) {$V_2$};

    \node (B3) at (2,4) {$B_3$};
    \node (B1) at (1,2) {$B_1$};
    \node (B2) at (3,2) {$B_2$};
    \node (V) at (2,0) {$V$};
    
    \node (F1) at (6,1) {$F_1$};
    \node (F2) at (6,-1) {$F_2$};
    \node (F3) at (8,0) {$F_3$};
    \node (V3) at (4,0) {$V_3$};
    
    \draw (T3) -- (T1);
    \draw (T3) -- (T2);
    \draw (T1) -- (V2);
    \draw (T2) -- (V2);
    
    \draw (B3) -- (B1);
    \draw (B3) -- (B2);
    \draw (B1) -- (V);
    \draw (B2) -- (V);
    
    \draw (F1) -- (V3);
    \draw (F2) -- (V3);
    \draw (F2) -- (F3);
    \draw (F1) -- (F3);

\end{tikzpicture}
    \label{fig:hobbies}
    \caption{Figure showing a graph clustered via 2CCED and 2CCEDVS (respectively)}

\end{figure}
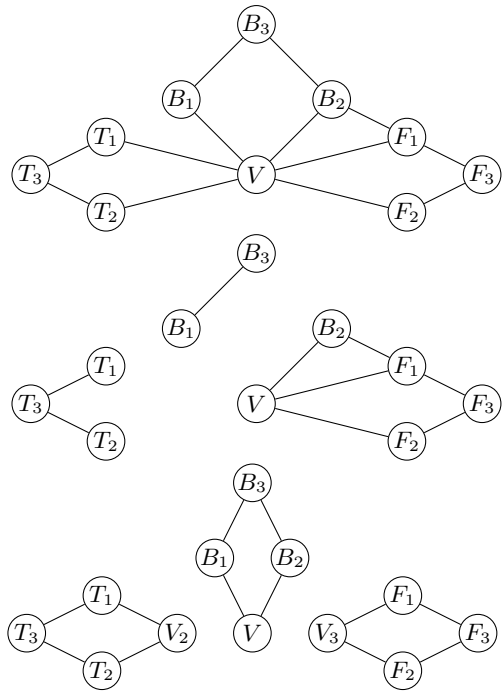

\vspace{5pt}
This problem was also shown to be $\NP$-Hard, but fixed-parameter tractable (FPT) when $s=2$ \cite{abukhzam2024complexity2clubclusterediting}. Figure \ref{fig:hobbies} below provides an example of the effectiveness of introducing the vertex splitting operation. 

\section{Heuristic Algorithms}

As mention above, {\sc $2$-Club Cluster Edge Deletion} (2CCED) and {\sc $2$-Club Cluster Edge Deletion with Vertex Splitting} (2CCEDVS) are both $\NP$-Hard. In practice, solutions implementing exhaustive search are too slow even when the input size is in the order of a thousand vertices.
In this section, we present two efficient heuristics for 2CCED and 2CCEDVS, which allows us to cluster larger graphs. 

The proposed 2CCED heuristic simply works as follows. At first, the square clustering coefficient is computed for each vertex in the graph. Next, a cluster is formed around the vertex with the highest square coefficient (i.e. the fraction of all possible squares that exist at a vertex $v$) such that all added vertices keep the cluster diameter to at most two (line 16). The final step is repeated until all vertices in the graph are in a cluster. The pseudocode of the 2CCED heuristic is presented below.

\begin{algorithm}[!htb]
\caption{$2$-Club Edge Deletion Heuristic}
\begin{algorithmic}[1]
\State $score \gets \{\}$ \Comment{Empty dictionary or hash map}
\State $clusters \gets []$ \Comment{Empty list to store clusters}
\For{each vertex $v$ in Graph}
    \State $score[v] \gets square\_coefficient[v]$ 
\EndFor
\State $score \gets reverse\_sort(score)$ \Comment{Sorting dictionary in descending order according to the score}

\State $unclustered \gets Vertices(Graph)$ 

\While{$unclustered$ is not empty}
    \For{each $v$ in $score$}
        \State $clusters[v] \gets v$
        \State $add\_s\_club(G, v, 2)$ \Comment{Add other elements}
        \State $unclustered \gets unclustered - clusters[v]$ 
    \EndFor
\EndWhile
\end{algorithmic}
\end{algorithm}

The algorithm does not require the input graph to have a similarity value (or another measure) between vertices. However, pairs of vertices with similarity below a certain threshold could be filtered out as a pre-processing step.

The 2CCEDVS heuristic works as follows.
For two non-assigned vertices having minimum or maximum degree $v$, we compute the transition matrix of the random walk on the graph $H_v$ whose vertex set is the closed $2$-neighborhood of $v$ and whose edge set consists of the induced edges of the graph $G$ and additional weighted arcs:
For every vertex $u$ at distance $2$ of $v$, we add an arc from $u$ to $v$ whose weight is the number of neighbors of $u$ not in the closed $2$-neighborhood of $v$.
In this random walk, the edges are considered to be 2 way arcs of weight $1$.

We multiply the $8^{th} $ power of this transition matrix by the vector consisting of zeroes except for $v$ where the value is $1$.
We remark that the $8^{th}$ power is enough to reach the stationary vector.
Otherwise, we could have replaced it by a norm to reach the difference between the vector and the iterated vector.
We order the vertices of the closed $2$-neighborhood by decreasing the probability computed by the latter vector.

For every prelist $X$ of this list which is a $2$-club, we compute the cost of this set of vertices as follows:
We compute for every vertex $x$ of $X$ the outdegree of $x$ which is the number of neighbors of $x$ not in $X$.
If the outdegree is $0$, then no operation will be operated for this vertex and the cost is $0$.
Otherwise, if the outdegree is $1$, then we will delete the edge and the cost is $1$.
Otherwise, the outdegree is greater than $1$ and in this case we will split the vertex so that we keep the neighbors in $X$ in one copy and we put the other neighbors (out of $X$) in the second copy.
The cost of $X$ is the sum of the costs of the vertices divided by the size of $X$.

Finally we keep the $2$-club cluster which minimizes the cost.
Then we iterate until all vertices are assigned to a cluster.
Our algorithm, presented below, can be adapted to consider similarity value between vertices.

\begin{algorithm}[!htb]
\caption{$2$-Club Edge Deletion with Vertex Splitting Heuristic}
\begin{algorithmic}[1]

\While{there exists an unassigned vertex}
    \State bestCluster $\gets \emptyset$
    \State $X \gets$ an unassigned min degree vertex
    \State add an unassigned max degree one to $X$
    \For{each $v$ in $X$}
        \State $A \gets$ transitionMatrix($v$)
        \State $l \gets $ order($v$, $A$)
        \State cluster $\gets$ bestPrefix($l$)
        \If{cost(cluster) $<$ cost(bestCluster)}
        \State bestCluster $\gets$ cluster
        \EndIf
    \EndFor
    \State process(bestCluster)
\EndWhile
\end{algorithmic}
\end{algorithm}

\section{Experiments}

We implemented our heuristics using Python. We did not compare against algorithms such as $k$-means and Hierarchical Clustering as they are not correlation clustering methods and require additional parameters such as the number of clusters. 

The results of MCL largely vary based on the choice of the inflation parameter whose typical range is [1.2-5.0] as per the MCL manual. For each dataset, we selected the optimal results across all tested inflation values, specifically 1.2, 1.3,..., 5.0.

All the experiments were conducted on a PC running Ubuntu 22.04.4 LTS with an Intel(R) Core(TM) i5-8350U CPU and 8GBs of RAM. The algorithms were tested on real and synthetic networks. For consistency, we disregarded any weights in the used networks, but we should note that our heuristics work on weighted networks and give similar results.

\subsection{Biological Networks}

The clustering algorithms were also tested on various Biological Networks from the Network Repository \cite{nr} whose properties are showcased in Table \ref{tab:properties}. These networks represent a gene-function association: vertices represent a gene or a function and an edge between a gene vertex and a function vertex imply that the gene has this specific function. Clustering this type of networks helps in understanding the organization of gene-sets, improving how they are interpreted \cite{yoon2019gscluster}.
For these graphs, the F-score cannot be computed since the ground-truth communities are not known. We therefore adopted the average intra-cluster and inter-cluster distances, which have been commonly used in the literature \cite{wu2004clustering,kumar2014performance,asur2007ensemble,zhai2020ad,halkidi2002clustering}. Our results are presented in Table \ref{tab:combined-results}.

\begin{table}[H]
\centering
\begin{tabular}{|c|c|c|c|}
\hline
Graph & Vertices & Edges & Clustering Coefficient \\ \hline
bio-CE-LC & 1.4k & 1.6k & 0.075 \\ 
bio-CE-HT & 2.6k & 3k & 0.008\\ 
bio-CE-GT & 924 & 3.2k & 0.605\\ 
bio-DM-HT & 3k & 4.7k & 0.009\\ \hline
\end{tabular}%
\vspace{2pt}
\caption{Properties of the different graphs used
}
\label{tab:properties}
\end{table}

\begin{table}[H]
    \centering
    \begin{tabular}{|c|c|c|c|c|}
    \hline
    Dataset & Algorithm & Time (s) & Intra-CD & Inter-CD \\ \hline
    bio-CE-LC & ClusterOne & 0.274 & 1.99 & 7.87 \\ 
    & MCL & 0.167 & 1.93 & \textbf{8.0} \\
    & 2CCED &{\textit{42.7}} & {\textit{1.92}} & \textbf{\textit{7.96}} \\ 
    & KaPoCE & 158 & {1.14} & 7.82 \\ 
    & 2CCEDVS & \textit{8.686} & \textbf{\textit{1.89}} & \textbf{\textit{7.59}} \\ \hline
    bio-DM-HT & ClusterOne & 0.631 & 2.24 & 8.10 \\ 
    & MCL & 0.501 & 1.69 & \textbf{8.66} \\ 
    & 2CCED & \textit{1.9} & \textbf{\textit{1.61}} & \textbf{\textit{8.14}} \\ 
    & KaPoCE & 390 & \textbf{1.18} & 8.14 \\ 
    & 2CCEDVS & \textit{124} & \textbf{\textit{1.56}} & \textbf{\textit{7.80}} \\ \hline
    bio-CE-HT & ClusterOne & 0.461 & 2.05 & 8.50 \\
    & MCL & 0.289 & 1.69 & \textbf{8.66} \\
    & 2CCED & \textit{1.86} & \textit{1.69} & \textbf{\textit{8.66}} \\
    & KaPoCE & 288 & \textbf{1.15} & 8.65 \\ 
    & 2CCEDVS & \textit{44} & \textit{1.62} & \textbf{\textit{8.37}} \\ \hline
    bio-CE-GT & ClusterOne & 0.269 & 2.06 & 3.77 \\
    & MCL & 0.272 & 1.82 & 3.77 \\
    & 2CCED & \textit{20.3} & \textbf{\textit{1.89}} & \textbf{\textit{3.78}} \\
    & KaPoCE & 210 & \textbf{1.14} & 3.74 \\ 
    & 2CCEDVS & \textit{720} & \textit{1.54} & \textit{3.24} \\
    \hline
    \end{tabular}
    \vspace{2pt}
    \caption{Combined Results for Clustering Algorithms across Different Datasets}
    \label{tab:combined-results}
\end{table}

\subsection{The Lancichinetti-Fortunato-Radicchi Benchmark}

We used an extended version of the Lancichinetti-Fortunato-Radicchi Benchmark (LFR) \cite{PhysRevE.78.046110} for undirected weighted overlapping networks\footnote{https://github.com/eXascaleInfolab/LFR-Benchmark\_UndirWeightOvp}. We generated multiple graphs by varying the parameter $p_{out}$ from $0$ to $0.12$ and the number of edges of vertices from 50 to 200. On average, the graphs contained 137.10 vertices and 1149.58 edges. To measure the clustering quality, we calculated the F-Score for the clusters, a standard practice when the ground-truth communities are known. The average results across all graphs are presented in Table \ref{tab:lfr-algorithm-comparison}.

\begin{table}[H]
\centering
\begin{tabular}{|l|c|c||c|c|c|c|}
\hline
\multirow{2}{*}{Algorithm} & \multirow{2}{*}{F-score} & \multirow{2}{*}{Time (s)} & \multicolumn{4}{c|}{Graph Properties} \\
\cline{4-7}
& & & Vertices & Edges & Clusters per Vertex & $p_{out}$ \\
\hline
ClusterOne & 0.8874 & \textbf{0.1916} & \multirow{6}{*}{137.10} & \multirow{6}{*}{1149.58} & \multirow{6}{*}{1.3044} & \multirow{6}{*}{0.0377} \\
MCL & 0.7789 & 1.5020 & & & & \\
2CCED & 0.6675 & 0.9830 & & & & \\
KaPoCE & 0.5959 & 36.0242 & & & & \\
2CCEDVS & \textit{\textbf{0.9159}} & \textit{4.9570} & & & & \\
\hline
\multicolumn{3}{|c||}{Range} & [50, 200] & [119, 2334] & [1.00, 1.93] & [0.0003, 0.1329] \\
\hline
\end{tabular}
\vspace{1pt}
\caption{Average Performance on the Extended LFR Benchmark}
\label{tab:lfr-algorithm-comparison}
\end{table}

\subsection{Analysis}

In the extended LFR Benchmark, 2CCEDVS had the best F-score with ClusterOne ranking second, and all other algorithms obtaining a significantly worse score. In general, when clusters overlap more, 2CCEDVS has the best combination of precision, recall, and accuracy.
In terms of execution speed, ClusterOne had the fastest average time. However, this came at the expense of clustering quality as it had the second to last F-score.
Benchmarks show that for the same program, C and C++ are faster than Java and Java is many times faster than Python \cite{fourment2008comparison,pereira2021ranking}. Using the same programming language would significantly reduce the gap in execution times among all three algorithms.  
Interestingly, KaPoCE had the worst executing time and clustering quality, and 2CCED significantly outperformed it in both measures. This confirms previous work indicating that relaxed models of cluster editing tend to result in better quality clusters.

In the biological data, KaPoCE had the best (smallest) intra-cluster distance across all graphs. This is expected as the objective of the heuristic is to obtain a disjoint union of almost-cliques. 2CCEDVS had the second best intra-cluster distances while the other algorithms fell behind. ClusterOne had the worst intra-cluster distance despite obtaining the second best F-score. At first glance, this might be surprising, but clustering quality metrics tend to negatively correlate with each other \cite{almeida2011there,emmons2016analysis}. In general, the inter-cluster distances were close to each other among all algorithms. Finding the best clustering quality when ground-truth communities are not known is still an ongoing area of research so obtaining the F-score when possible is still deemed to be the most reliable measure. 
We did not use the intra-cluster and inter-cluster measures for LFR since the F-score is known to be the most indicative. Yet, we should probably note that our algorithms are also competitive with respect to these measures: 2CCED exhibited the best intra-distance measure after the {\sc Cluster Editing} algorithm while 2CCEDVS ranked second after MCL with respect to the inter-cluster measure.

One main downside of MCL is the need to perform hyper-parameter tuning in order to obtain clusters with good quality. Although some options are available to make the process easier, they only target a few quality measures. The other algorithms did not require any hyperparameter tuning or pre-processing steps to obtain the presented results. In practice, the extra steps needed to obtain good quality clusters with MCL makes the process slower and less user-friendly.

\section{Conclusion}

We studied a new model for correlation clustering that is based on graph modification by combining edge-deletion into 2-clubs and vertex splitting. In addition to being a reasonable relaxed version of correlation clustering, our model permits a more realistic ``grouping'' of data elements. In fact, by allowing a data element to be cloned via splitting, and combining this with our distance-based measure of closeness, we can better compute overlapping clusters. 

We presented experimental results for two data categories that cover both synthetic and real data. The results clearly showcase the ability of our proposed method to obtain better clustering. 
We should note that outperforming all the tested methods on the LFR benchmark was a surprising outcome that we have not anticipated. This calls for further research and scrutiny. 

The number of times a vertex splits can  be used as another parameter. In future work, it would be interesting to study a multi-parameterized version of the problem, where the maximum number of operations performed per single vertex is controlled by input parameters, as in \cite{abu2017complexity}. This approach proved to be useful in a number of application domains \cite{barr2019combinatorial,barr2020vulnerability,anomaly} where it is unlikely for a data element to belong to an arbitrary number of communities.

The approach introduced in this paper is heuristic. This is partially affected by the fact that the only known exact (fixed-parameter) algorithm presented in \cite{abukhzam2024complexity2clubclusterediting} is far from practical. It would be interesting to combine the implicit backtracking method of \cite{abu2010} with an improved fixed-parameter algorithm that runs in $\bigo^*(c^k)$, for a small constant $c$. This can possibly result in further improved clustering. However, efficiency remains an issue on large data sets. In fact, even our heuristic method suffers from lack of efficiency on large data sets, mainly because 
we have a number of intermediate computations that are based on graph traversal and matrix computations.
In future endeavors, we plan to optimize the speed of the presented heuristic and implement a version that runs on GPUs in order to cluster large networks. Other directions that can be explored include the selection of the best value of $s$ in $s$CCEDVS, and possibly some problem variants with local parameters bounding the number of  modification operations for each vertex.

\end{document}